\documentclass[aip,jcp,amsmath, amssymb,reprint]{revtex4-1}

\usepackage{amsmath,amssymb}
\usepackage{graphicx}
\usepackage{xcolor}

\newcommand{\mydef}
        {\stackrel{\mathrm{def}}{=}}

\newcommand{\Oone}{\mathrm{O}}
\newcommand{\Otwo}{\mathrm{O}_2}
\newcommand{\COtwo}{\mathrm{CO}_2}
\newcommand{\CO}{\mathrm{CO}}

\newcommand{\ad}{\,\mathrm{(ad)}}

\newcommand{\todo}[1]{\noindent\textcolor{red}{$\Box$ #1}}

\begin{document}

\title{Detailed analysis of transitions in the CO oxidation on Palladium(111) under noisy conditions}

\author{Jaime Cisternas}
\email{jecisternas@miuandes.cl}
\affiliation{Complex Systems Group, Facultad de Ingenier\'{\i}a y Ciencias Aplicadas,
Universidad de los Andes, Monse\~nor Alvaro del Portillo 12455, Las Condes, Santiago, Chile}
%
%
%
\author{Stefan Wehner}
\affiliation{Surface Science Group, Institute for Integrated Natural Sciences, University of Koblenz-Landau, Universit{\"a}tsstra{\ss}e 1, 56070 Koblenz, Germany}

\date{\today}

\begin{abstract}
It has been shown that $\CO$ oxidation on Pd(111) under ultra-high vacuum conditions can suffer rare transitions between two stable states triggered by weak intrinsic perturbations.
Here we study the effects of adding controlled noise by varying the concentrations of $\Otwo$ and $\CO$ that feed the vacuum chamber, while the total flux stays constant.
In addition to the regime of rare transitions between states of different $\COtwo$ reaction rate induced by intrinsic fluctuations,
we found three distinct effects of external noise depending on its strength:
small noise suppresses transitions and stabilizes the upper rate state;
medium noise induces bursting; and
large noise gives rise to reversible transitions in both directions.
To explain some of the features present in the dynamics,
we propose an extended stochastic model that includes a global coupling through the gas phase to
account for the removal of $\CO$ gas caused by the adsorption of the Pd surface.
The numerical simulations based in the model show a qualitative agreement with the noise-induced
transitions found in experiments, but suggest that more complex spatial phenomena are present in
the observed fluctuations.


\end{abstract}

\maketitle


\section{Introduction}

Surface reactions provide a convenient experimental realization of a
pattern forming system outside of equilibrium. Particularly in the case of
CO oxidation on
Platinum-group single crystals, it has been possible to implement effective
techniques to control the conditions of the reaction and measure a variety of spatio-temporal behaviors.
In parallel to these experimental developments, a clear theoretical picture has
been established, connecting the microscopic behavior of molecules with the
nonlinear effects at the mesoscopic scale, often modeled by reaction-diffusion equations.

Among the experimental breakthroughs one should highlight the ultra-high-vacuum
techniques that allow precise control of the elementary processes by manipulation
of temperature and partial pressures. With these techniques it
was possible to explore the conditions where the mesoscopic models predict instabilities.
At these critical points (bifurcations) the homogeneous and stable coverages of the surface become unstable
and new spatial patterns arise, often accompanied with oscillatory or more
complex temporal dependence.
Close to the conditions of instability, not just external perturbations can be magnified
by the reaction system, but even microscopic fluctuations or surface defects can become relevant for the overall dynamics.

In this work we consider the effects of controllable noise on the CO oxidation on Pd(111),
in a regime where two different stable states coexist (as measured by $\COtwo$ reaction rate), but relatively close to the point
where one of them becomes unstable, as explained in Ref.~\onlinecite{KWK09}.
Previous work about noise and $\CO$ oxidation on other surfaces is documented in Refs.~\onlinecite{PT09,BAR09,FXXWK17}.
This Palladium surface, that does not have reconstruction and does not show oscillations,
can exhibit a behavior that is not observable in other Platinum-group crystals.
When only intrinsic noise (due to experimental limitations) is present and
for very specific control parameters inside the bistable range,
we have observed rare reversible transitions between the two stable states.
Residence times in these two states can go from 5000 seconds to 10000 seconds and more,
see Ref.~\onlinecite{WKBSKB10} for more details including spatially resolved experiments by photo-electron emission microscopy (PEEM).
A earlier study of the authors explained this observation, see Ref.~\onlinecite{CKW14} for details.

The main motivation of this work was to enhance the occurrence of reversible transitions by adding
homogeneous noise, assessing the most effective combinations of distance to the critical point
and noise intensity. Our results indicate that on the one hand
small noise 
can have a surprising suppressing effect for
the rare transitions, but on the other hand large noise induces reversible
transitions between the stable states.
This work continues the research by the authors on noise-induced transitions on other surfaces, mostly Iridium(111) \cite{HWKB04,WHBK04,WHSBK05,HWSBK06,WHSBK06,CEDW09,CEDW10,CLW11,CWD12,WCDK14}.

In this article we first present the experimental setup and the results obtained for
a variety of distances to the critical point and noise intensities.
Several behaviors are studied in detail, highlighting key features that will be
incorporated in the modeling, notably the global effect of adsorption on the available concentrations of gases inside the vacuum chamber.
The second part of the article is devoted to the model based on partial differential equations, that includes global coupling as
two additional ordinary differential equations.
Although the numerical results for the model are restricted to a one-dimensional surface,
there is a close qualitative agreement with the experiments.
We conclude the article with a general assessment of our experimental findings
and the successes and shortcomings of the model, that prompt more extensive studies
of the phenomena associated with rare and induced transitions.

\section{Experimental setup and results}


For a detailed study of the elementary steps and nonlinear effects of the conversion from $\CO$ to $\COtwo$ on a catalyst, ultra-high-vacuum conditions are needed. Here the (111) surface of a Palladium single crystal was used as the catalyst for conversion. The Pd
crystal has a polished face of about $1\ \mathrm{cm^2}$ and was located in an UHV system consisting of a small vacuum vessel and a large one. Both are connected by a circular orifice (9 mm diameter), which is closeable by a shutter mechanism. Via a capillary in the small vacuum system the feed gas is delivered from a gas system based on two mass flow controllers (MFCs, MKS 200) in a composition that can be selected dynamically. The flux of the feed gas is fixed through the present study at 1 ML/s.
Both MFCs are controlled by a computer-generated signal $Y(t)$ between 0 and 1 that specifies the molar fraction of CO.
The control signal $Y(t)$ is a piecewise constant function of characteristic time $t_\mathrm{step}$ (set to 3 s throughout this work, see Ref.~\onlinecite{WCDK14} for details), average $Y_0$ and standard deviation $\Delta Y/2$.
Each part of the system is pumped by stacked turbomolecular pumps, and therefore the experimental setup represents a continuously pumped reactor. For analysis a quadrupole mass spectrometer (Balzers QMS 421) is located in the small chamber. Since the pumping speed is adequately set, the $\COtwo$ partial pressure (amu 44) recorded by the QMS is proportional to the $\COtwo$ reaction rate. Measuring a signal of $x$ amu means that the QMS detected a signal for m/e-ration of $x$ amu.
Additionally some $\COtwo$ molecules experience fragmentation during the ionization process inside the QMS. This fragmentation pattern is specific for each setup. 
The fragmentation pattern of $\COtwo$ molecules contains also a fragment with 28 amu.

On the other hand amu 28 is also the mass of $\CO$. Therefore the signal of amu 28 has to be handled with more care to ensure that it is a proper measure for the $\CO$ content in the chamber. With decreasing $\COtwo$ reaction rate on one hand the amu 28 signal will decrease due to cracking of $\COtwo$ molecules, but on the other hand it will increase since the number of $\CO$ molecules rises due to lack of consumption via oxidation on the surface. Since the partial pressure of $\COtwo$ was typically one to two orders of magnitude smaller than the back pressure of $\CO$ in the chamber, the contribution to amu 28 from $\COtwo$ can be neglected and the QMS signal amu 28 can be used in this study as partial pressure of $\CO$ in the chamber i.e. the environment of the Palladium surface.

\begin{figure}
\begin{center}
\includegraphics[width=0.5\textwidth]{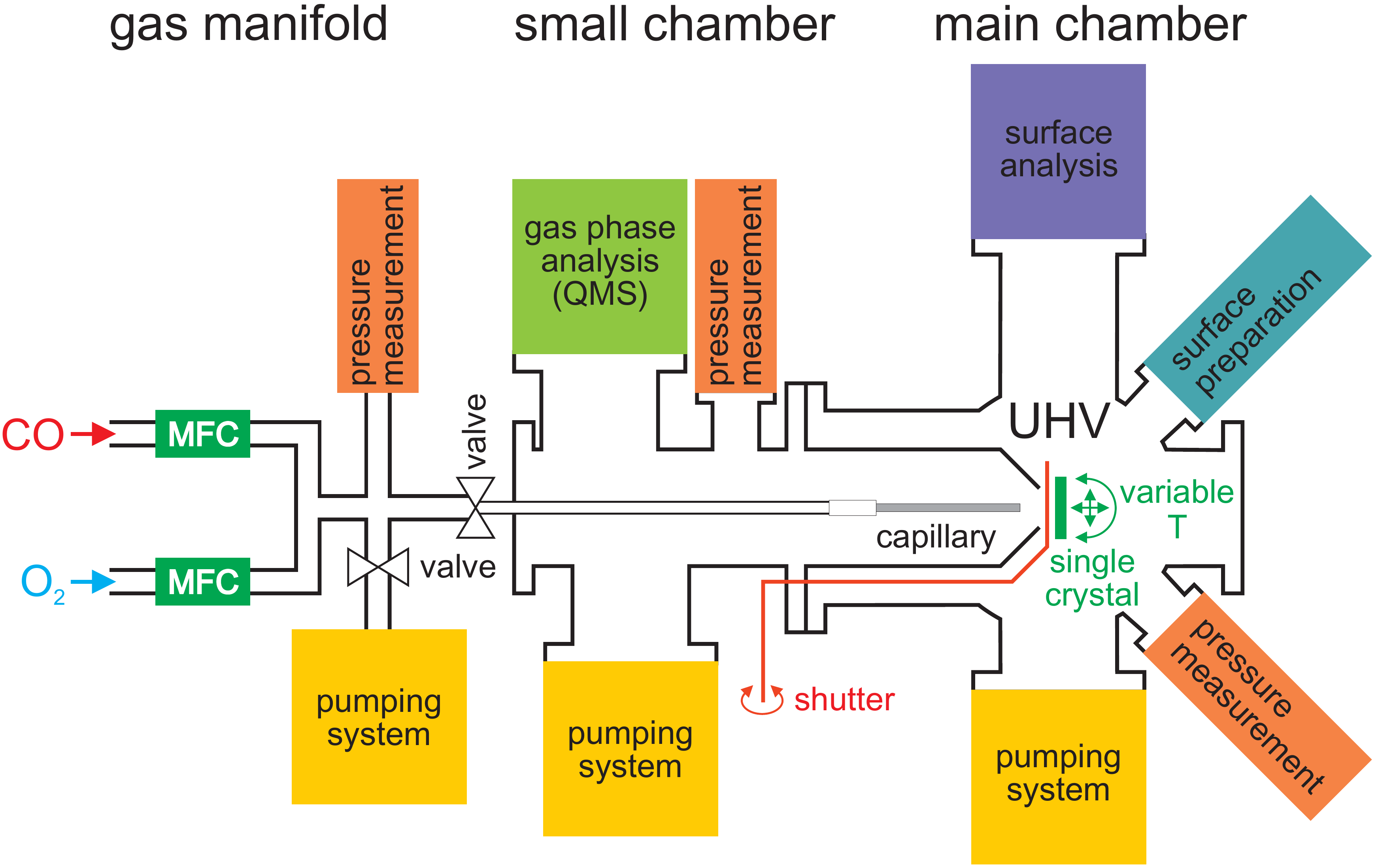}
\bigskip
\end{center}
\caption{Schematic of the experimental setup where ultra high vacuum conditions are used. From left to right: gas manifold with two mass flow controllers (MFC) dosing the feed gas via a capillary inside the small chamber, the single crystal surface is located in side the main chamber behind a closeable orifice. The reaction takes place on the heated surface, the partial pressure of gases are measured with a quadrupole mass spectrometer (QMS).}
\label{fig:setup}
\end{figure}

The large chamber hosts standard UHV cleaning and surface characterization equipment. For measurements the crystal is located in front of the open orifice so the surface where the reaction takes place is directly exposed to the feed gas. The $\COtwo$ rate recorded in the small vessel with a QMS is summing up contributions from all reactions taking place on the whole surface. 
The schematic of the experimental setup is shown in Fig.~\ref{fig:setup}. A more detailed description of the experimental setups can be found in our previous studies \cite{KWK09, WKBSKB10, CKW14}.
More details on the methods of measurement were published in the reviews Refs.~\onlinecite{W09, WCDK14}. It should be emphasized that `$\CO$ set' refers to the condition after the MFC; `$\CO$ feed' refers to the condition after smoothing by the capillary; and `$\CO$ in chamber' to the partial pressure of $\CO$ as recorded by the QMS. \cite{CWD12, WCDK14}
This nomenclature will be used throughout the present work, particularly when presenting experimental results.

The surface reaction is described by the Langmuir-Hinshelwood mechanism:
\begin{equation*}
2\ \CO + \Otwo \rightarrow 2\ \COtwo ~,
\end{equation*}
that involves the following elementary processes:
adsorption of both reactants, molecular for $\CO$ and dissociative for $\Otwo$:
\begin{align*}
\CO \rightarrow \CO\ad ~, \\
\Otwo \rightarrow 2\ \Oone\ad ~;
\end{align*}
desorption of adsorbates (because of the chosen surface temperature 410 K only $\CO$ desorption has to be included):
\begin{equation*}
\CO\ad \rightarrow \CO ~;
\end{equation*}
and reaction, when $\CO$(ad) and $\Oone$(ad) meet while diffusing on the surface:
\begin{equation*}
\CO\ad + \Oone\ad \rightarrow \COtwo ~.
\end{equation*}

When the CO content in the feed gas (controlled by the MFCs) quantified by the molar fraction $Y(t)$ is slowly varied cyclically, see Ref.~\onlinecite{KWK09}, a hysteresis is observed in the $\COtwo$ rate for a limited region in molar fraction $Y$ and temperature $T$. Two branches can be observed:

\begin{itemize}
\item[UR:] upper $\COtwo$ rate, O covered surface;
\item[LR:] lower $\COtwo$ rate, CO covered surface. 
\end{itemize}

The UR stable state exists for values of molar fraction $Y$ between $0$ and $Y_h$; and
the LR stable state exists for values of molar fraction $Y$ between $Y_l$ (smaller than $Y_h$) and 1.
The bistability range is defined by $Y_l < Y < Y_h$.
The limits $Y_l$ and $Y_h$ depend basically on temperature and feed flux.
 

\begin{figure*}
\begin{center}
\includegraphics[width=1\textwidth]{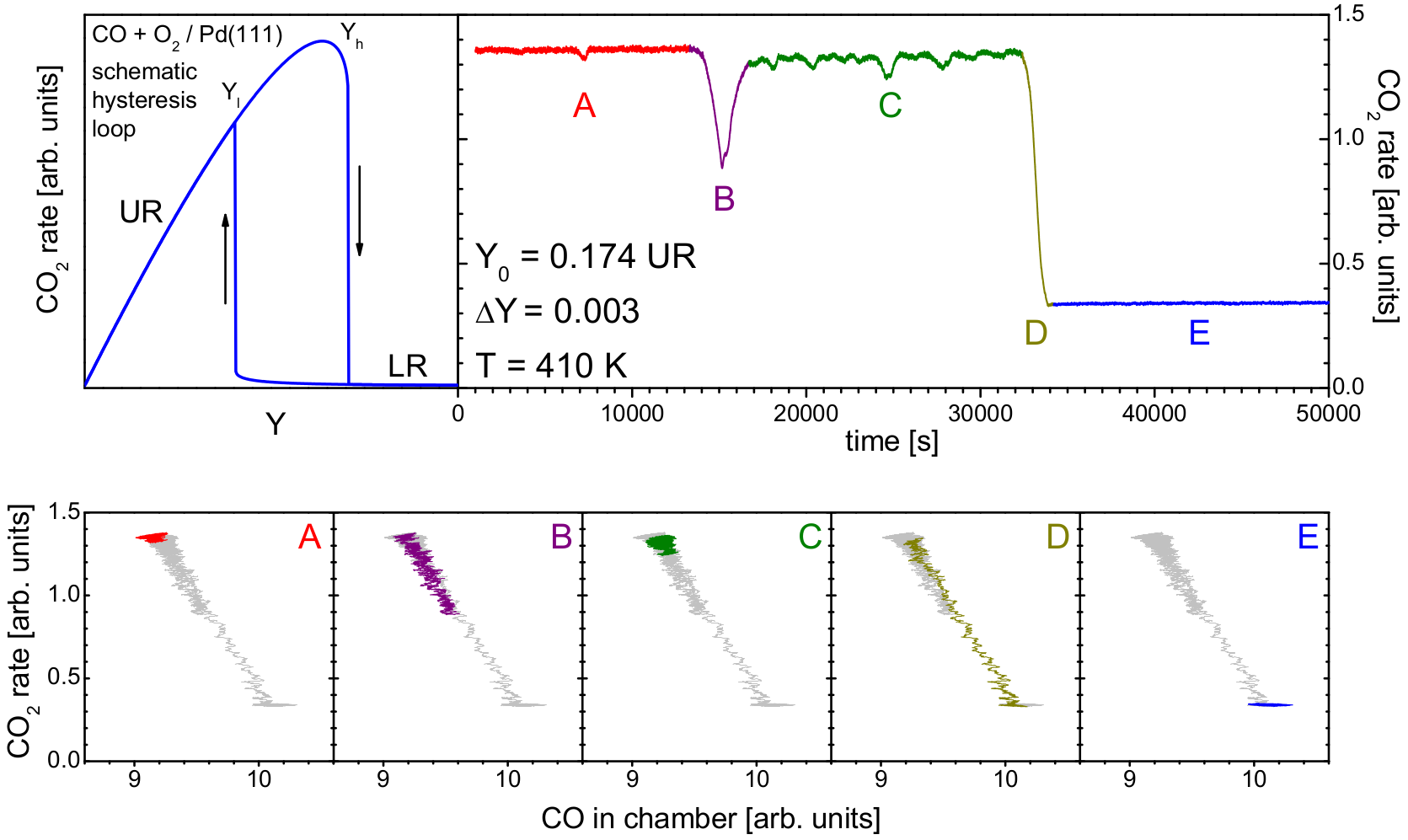}
\bigskip
\end{center}
\caption{
Upper panel left: Schematic hysteresis loop for $\CO$ oxidation on Pd(111).
Upper panel right: An exemplary experimental run for $\CO$ oxidation on Pd(111) at 410 K. The surface was prepared in upper rate (UR) for $Y_0 = 0.174$ and recorded for 50000 s, superimposing weak noise of strength $\Delta Y = 0.003$. The $\COtwo$ rate is shown as function of time. Characteristic regions are color coded and labeled A to E showing spontaneous $\CO$ poisoning.
Lower panel: For each characteristic region A to E the $\COtwo$ rate vs. $\CO$ in chamber is shown. The light grey line shows the total data of the above experiments, and each one of the colored data represents one of the characteristic regions.
}
\label{fig:hysteresis}
\end{figure*}

Based on experience one could expect that inside the bistable range, the final state depends on the initial preparation of the surface, predominantly $\CO$ resp. $\Oone$ covered. For Ir(111) without external perturbations the state of the system stayed at one of the fixed states and most of the recorded fluctuations were small. Adding small to medium noise, transitions towards the \emph{globally} stable state were observed, and adding large noise, back-and-forth bursts and switchings between both rates were recorded. This general behavior was observed for the $\CO$ oxidation on Ir(111), details can be found in Refs.~\onlinecite{WBK03, WBRK03, WHSBK05, HWKB04, HWSBK06, WCDK14, W09}.


In experiments with $\CO$ oxidation on Pd(111) recorded for a long time a transition from UR state to LR state is observed. The transition is shown in Fig.~\ref{fig:hysteresis} (upper panel), where the $\COtwo$ rate drops at around $t=33000$ s to a very low signal where there is no recovery. Only a period of at least one minute with pure oxygen (procedure to prepare a UR) can overcome this state. Within the experiment under reacting conditions these transitions are irreversible.

In the upper right panel of Fig.~\ref{fig:hysteresis} an experimental run of the $\CO$ oxidation on Pd(111) at 410 K is shown. Initially the surface is cleaned with an oxygen pulse and the appropriate CO content is selected, here $Y_0 = 0.174$. 
The surface is therefore prepared in upper rate (UR) for $Y_0 = 0.174$ and during the measurement (here 50000 s) superimposed with noise of intensity $\Delta Y = 0.003$ and characteristic time $t_\mathrm{step} = 3$ s. 
The $\COtwo$ rate is color coded for five characteristic regimes labeled A to E. 
In the bottom row, and for each of these five sections, the $\COtwo$ rate is shown versus the partial pressure of $\CO$ measured inside the chamber. 

Section A (red) shows the regime of noisy UR, exhibiting only small fluctuations. In this state the highest $\COtwo$ rates were recorded.
Section B (purple) contains a spontaneous and reversible transition, similar to those addressed in Refs.~\onlinecite{WKBSKB10, CKW14}. During this incomplete transition the $\COtwo$ rate decreases to a value as low as about half of the former upper rate.
Section C (green) shows the recovered upper rate, where the observed fluctuations are larger and the frequency of small dips is increased. The $\COtwo$ rate is lower in average and less constant, see larger explored region in the lower panel.
Section D (dark yellow) represents the transition to the state \emph{poisoned} with CO. Compared to section B, the $\COtwo$ rate reaches its lowest values quite fast (the trajectory can be seen in lower panel). 
Section E (blue) represents the noisy LR. $\COtwo$ rate fluctuations are now quite small, even when noise is still present.
The lower panels of Fig.~\ref{fig:hysteresis} indicate a characteristic linear pattern 
between the increase of CO inside the chamber and the decrease of $\COtwo$ rate.

\begin{figure}
\begin{center}
\includegraphics[width=0.5\textwidth]{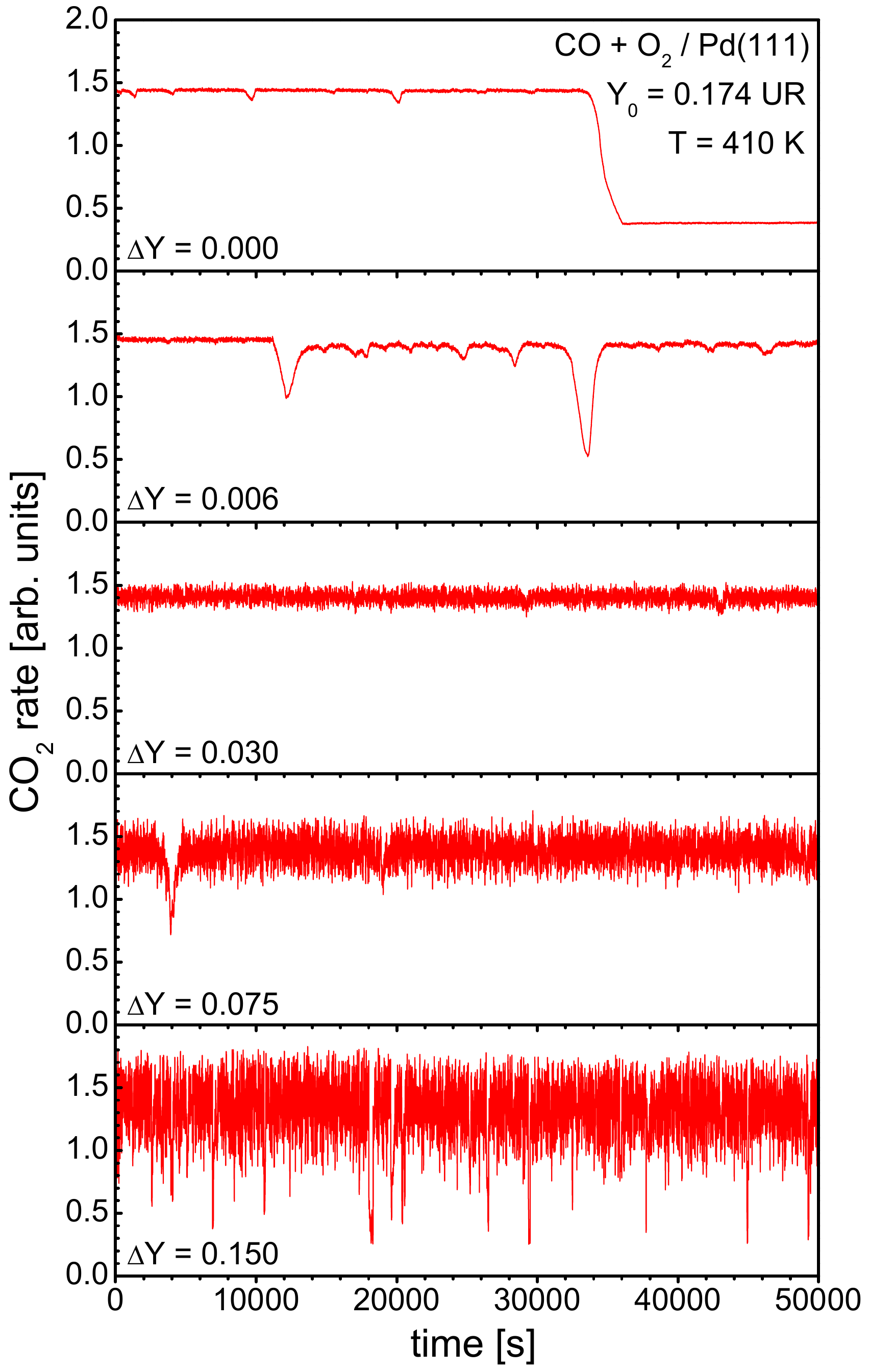}
\bigskip
\end{center}
\caption{Summary of observed behaviors for $Y_0=0.174$, for different levels of noise intensity $\Delta Y$:
the first two panels show the effects of intrinsic noise; and the last three show the effects of
small noise, medium noise, and large noise.
In each plot, the time evolution of the $\COtwo$ reaction rate is shown.}
\label{fig:Y174}
\end{figure}

In Fig.~\ref{fig:Y174} the influence of noise on the recorded $\COtwo$ rates in $\CO$ oxidation on Pd(111) at a surface temperature of 410 K and average molar fraction $Y_0 = 0.174$ is shown. Each panel shows the $\COtwo$ rate as a function of time for individual runs where the surface was initially prepared in upper rate. Starting from top to bottom the intensity of the superimposed noise is increased from $\Delta Y = 0.000$ (intrinsic) to $\Delta Y = 0.150$. 
With only intrinsic noise, a behavior similar to the one shown in sections A, D and E in Fig.~\ref{fig:hysteresis} is observed. For $\Delta Y = 0.006$, noisy upper rate (as in section A in Fig.~\ref{fig:hysteresis}) is seen first, after a first dip (as in section B in Fig.~\ref{fig:hysteresis}) the rate appears as in section C in Fig.~\ref{fig:hysteresis}, after another dip the rate recovers to a upper rate, somehow smoother than before, but still similar to section C in Fig.~\ref{fig:hysteresis}. No transition to a CO poisoned state is observed within a time window of 50000 s. 
For small noise $\Delta Y = 0.030$ the behavior is even more constant. Only small dips of $\COtwo$ rate could be seen. 
For medium noise $\Delta Y = 0.075$ a more emphasized dip is recorded, but in general no changes in the overall behavior are observed.
For large noise $\Delta Y = 0.150$ some reversible excursions to the lower rate are recorded. Such bursts and switchings between two stable states were already found for $\CO$ oxidation on Ir(111),
but on Pd(111) increasing noise seems to \emph{stabilize} the $\COtwo$ rates and to reduce the appearance of irreversible transitions. In contrast, on Ir(111) the transition towards the globally stable state was more frequent with increasing noise \cite{HWKB04}.

\begin{figure}
\begin{center}
\includegraphics[width=0.5\textwidth]{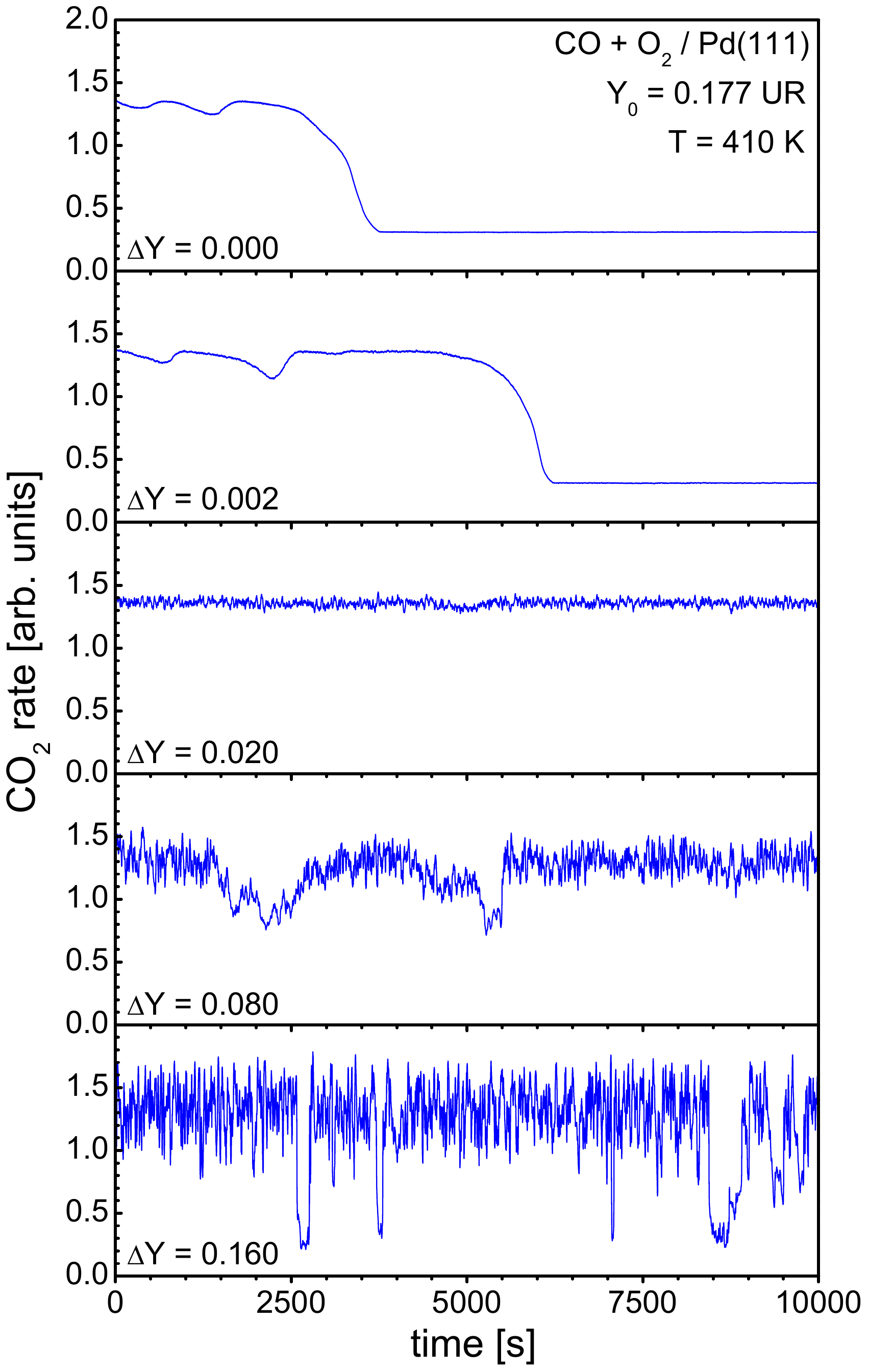}
\bigskip
\end{center}
\caption{Summary of observed behaviors for $Y_0=0.177$, for different levels of noise intensity $\Delta Y$.}
\label{fig:Y177}
\end{figure}

Figure \ref{fig:Y177} shows experiments performed using an average molar fraction $Y_0 = 0.177$, more distant to the point $Y_h$ where UR becomes unstable. 
The main difference with the previous experimental set ($Y_0=0.174$) is the shorter time per experiment, here 10000 s. Starting from top to bottom the superimposed noise is increased from $\Delta Y = 0.000$ (intrinsic) to $\Delta Y = 0.160$. 
As seen before, $\COtwo$ rates from experiments with intrinsic or weak noise ($\Delta Y = 0.002$) show a noisy upper rate with dips, a fast transition at 2500 s resp. 5500 s and finally the poisoned state. 
Small noise $\Delta Y = 0.020$ is able to stabilize the upper rate for the whole duration of the experiment, and no transition takes place.
For medium noise $\Delta Y = 0.080$ and large noise $\Delta Y = 0.160$ the general behaviors are similar to those in Fig.~\ref{fig:Y174}. $\COtwo$ rate exhibits bursts and switching phenomena, without significant changes in fluctuation level before and after the dips. 


From PEEM experiments and simulations of the $\CO$ oxidation on Ir(111) it is known that the transition towards the globally stable state (rate) is ruled by the growth of adsorbate islands on a mesoscopic scale ($\mu$m). On Ir(111), the growth of islands was not influenced by noise strength but only by the average $Y_0$ \cite{HWSBK06}. 
The transition time was found to decrease due to an increased number of adsorbate islands and not the faster growth of an individual island\cite{WHSBK05, HWSBK06}. 

For the $\CO$ oxidation on Pd(111), transitions seem to be hampered by noise.
Similar phenomena has been found in other contexts e.g. Ref.~\onlinecite{H08}.
One of the hypotheses is a larger threshold size of nucleated islands to grow on Pd(111). Therefore further research is needed, including spatially resolved experiments. e.g. PEEM
or ellipsomicroscopy for surface imaging (EMSI, as implemented for instance in Ref.~\onlinecite{PWKRE02}).

\begin{figure}
\begin{center}
\includegraphics[width=0.5\textwidth]{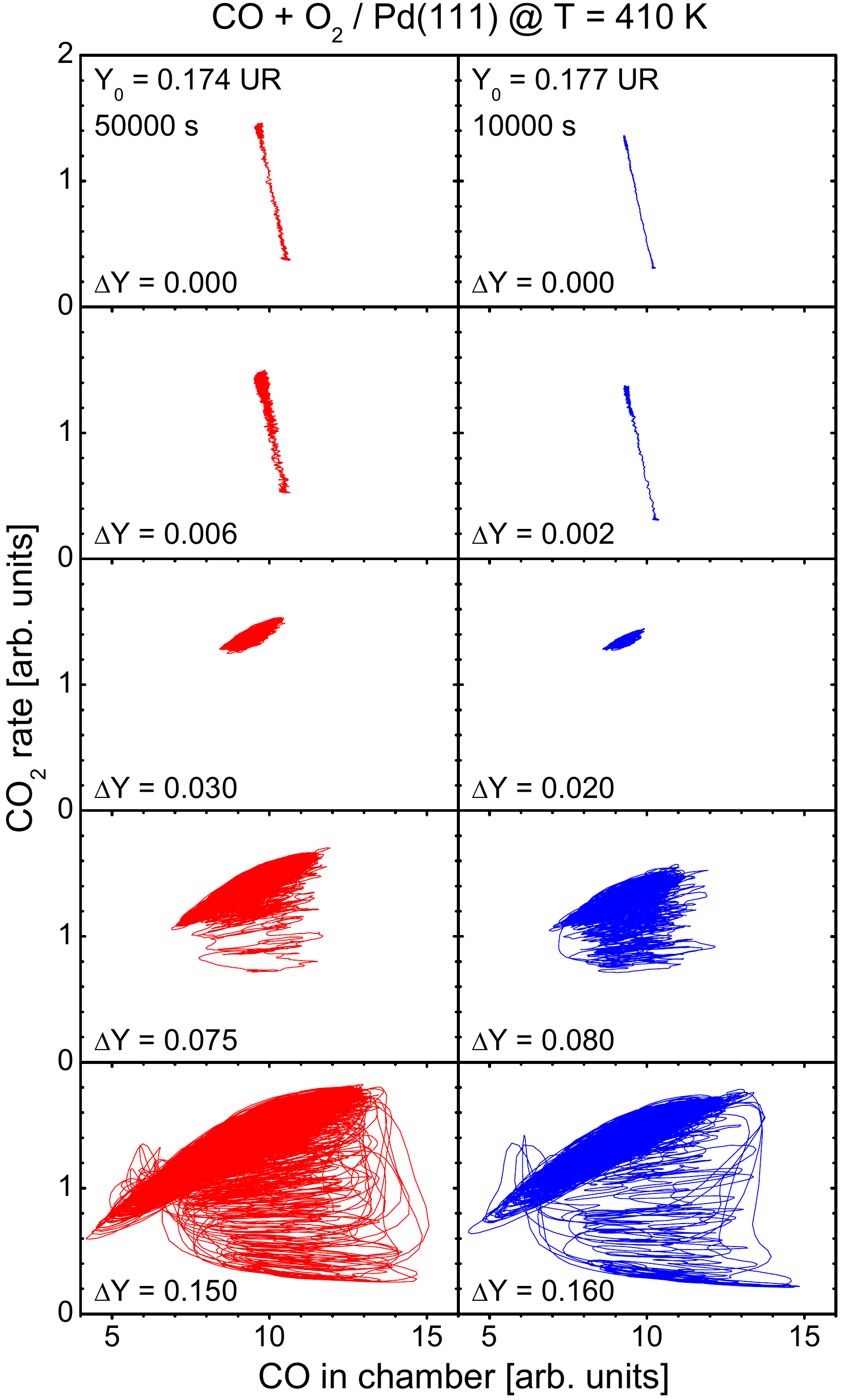}
\bigskip
\end{center}
\caption{$\COtwo$ rates from figures \ref{fig:Y174} (left column) and \ref{fig:Y177} (right column) versus gas $\CO$ concentration inside the chamber.}
\label{fig:overview}
\end{figure}

In the left and right columns of Fig.~\ref{fig:overview}, the data from Figs.~\ref{fig:Y174} and \ref{fig:Y177} are presented, using $\COtwo$ vs. CO in chamber (as in the lower panel of Fig.~\ref{fig:hysteresis}).
In this representation it is obvious, that when intrinsic noise dominates (two topmost rows) the system exhibits transitions from upper to lower rate (from higher left to lower right in each graph). 
For small noise (middle row) these transitions do not take place and the trajectories are restricted to compact areas at the upper left region of the graphs.
With further increases of noise intensity more of the phase space is explored. For medium noise (second lowest row) the trajectories do not reach the lower rate.
While for large noise (bottom row) nearly the whole area of the graph is visited by the trajectory, generating messy figures where it is still possible to recognize a hysteresis loop with well defined UR and LR branches. 
The two graphs of the last row also show `overshooting' as the state goes up from the LR (always in the clockwise direction) and reaches momentarily a $\COtwo$ rate higher than the UR branch.

\begin{figure}
\begin{center}
\includegraphics[width=0.5\textwidth]{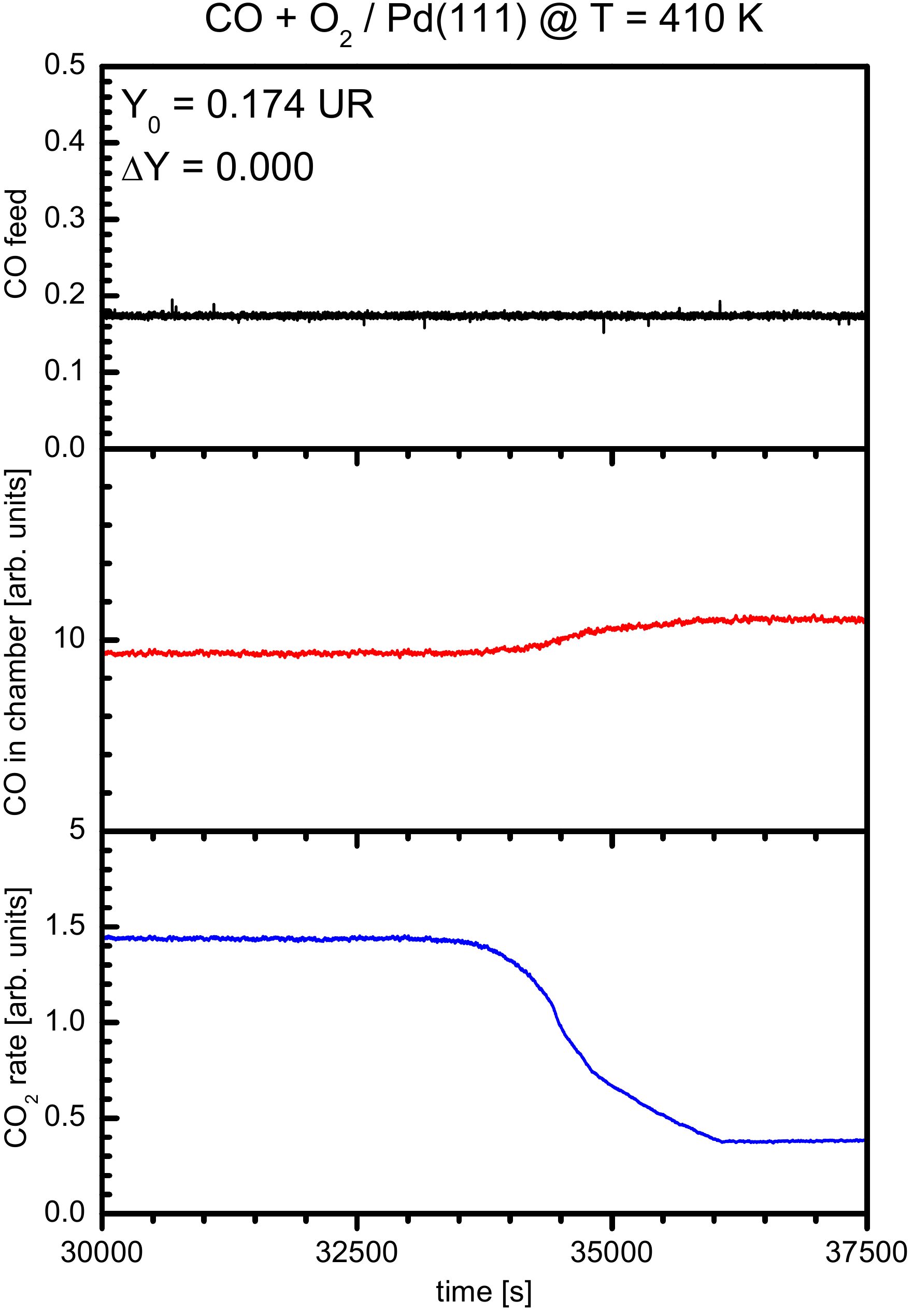}
\bigskip
\end{center}
\caption{CO feed (MFC reading), CO in chamber (QMS amu 28 signal) and $\COtwo$ rate (QMS amu 44 signal)
for a characteristic period of the experiment shown in Fig.~\ref{fig:Y174} for $Y_0 = 0.174$ and $\Delta Y = 0.000$ (intrinsic noise).
This figure illustrates a spontaneous transition (not driven by CO feed) and the characteristic enhancement of CO inside the chamber as a result of
the reduced adsorption of CO in the LR state.}
\label{fig:Y174_dY0000}
\end{figure}

\begin{figure}
\begin{center}
\includegraphics[width=0.5\textwidth]{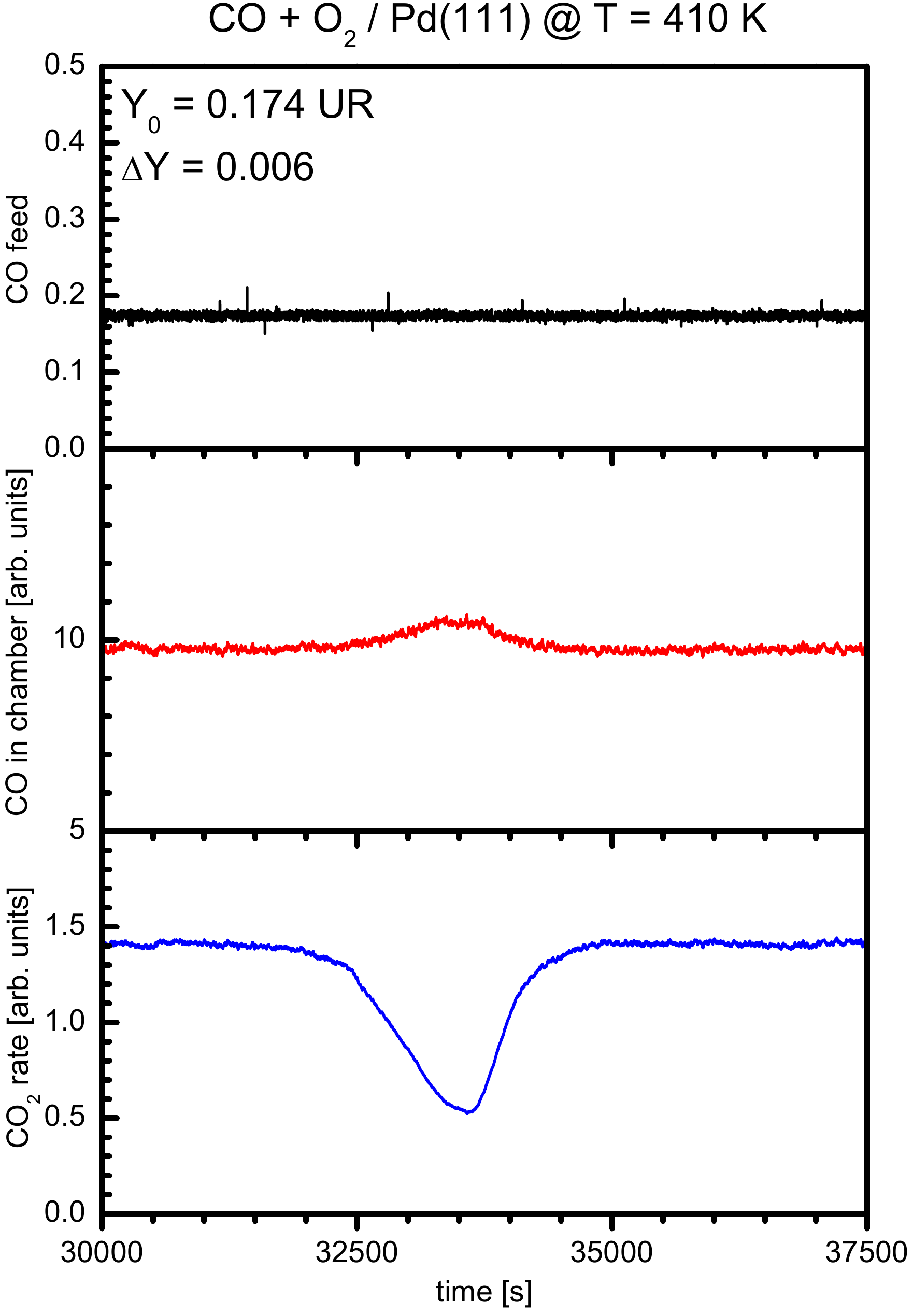}
\bigskip
\end{center}
\caption{As in figure \ref{fig:Y174_dY0000}, but for $\Delta Y = 0.006$.
This figure illustrates a spontaneous (not driven by CO feed) but incomplete transition and the characteristic enhancement of CO inside the chamber as a result of
the reduced adsorption of CO.}
\label{fig:Y174_dY0006}
\end{figure}

\begin{figure}
\begin{center}
\includegraphics[width=0.5\textwidth]{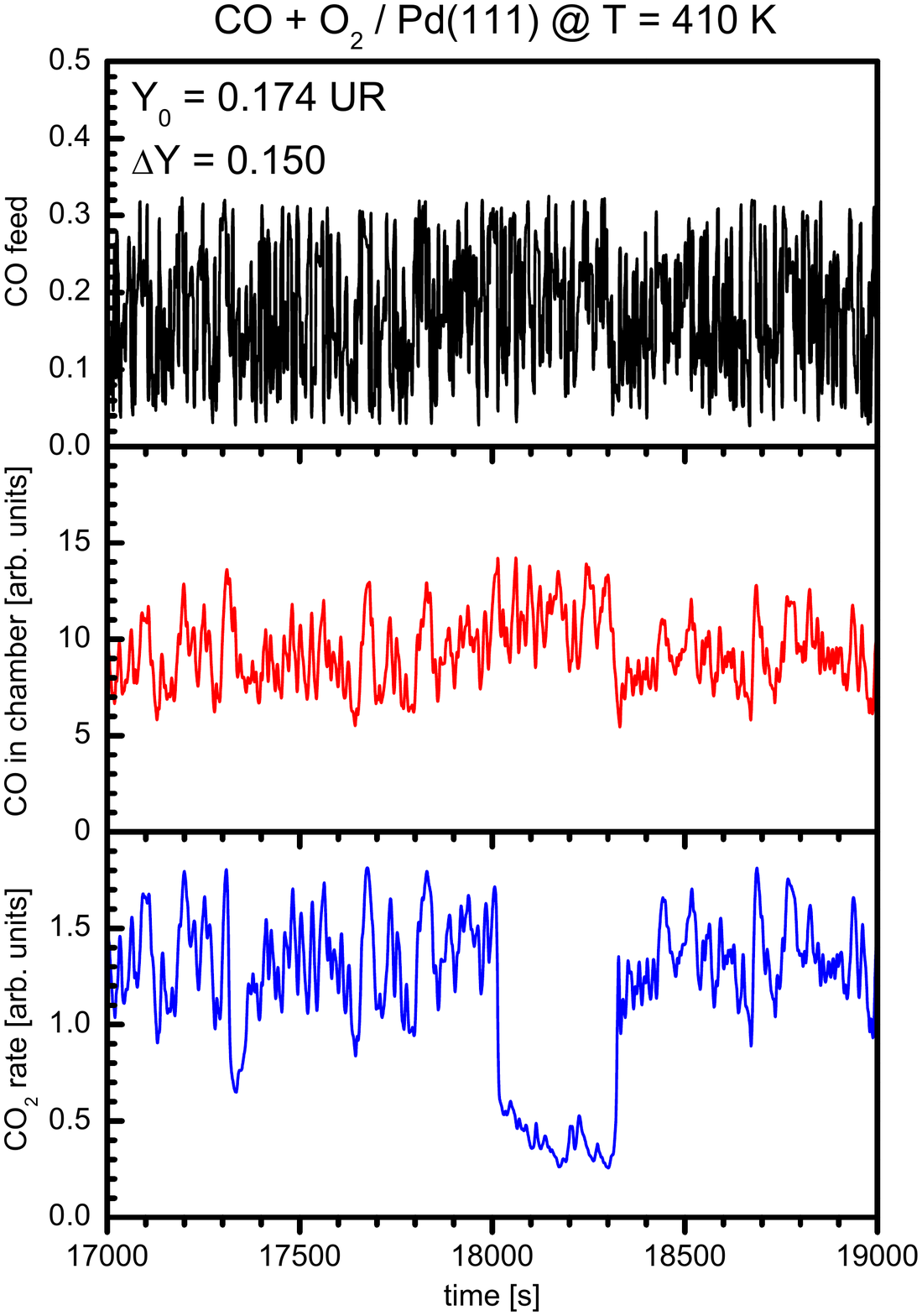}
\bigskip
\end{center}
\caption{As in figure \ref{fig:Y174_dY0000}, but for $\Delta Y = 0.150$.
This figure illustrates two noise-induced transitions (from UR to LR and its reversal) that are triggered when the CO concentration inside the chamber crosses certain threshold.
Also visible are the different in-phase and anti-phase fluctuation patterns of the UR and LR states (respectively).}
\label{fig:Y174_dY0150}
\end{figure}

\begin{figure}
\begin{center}
\includegraphics[width=0.5\textwidth]{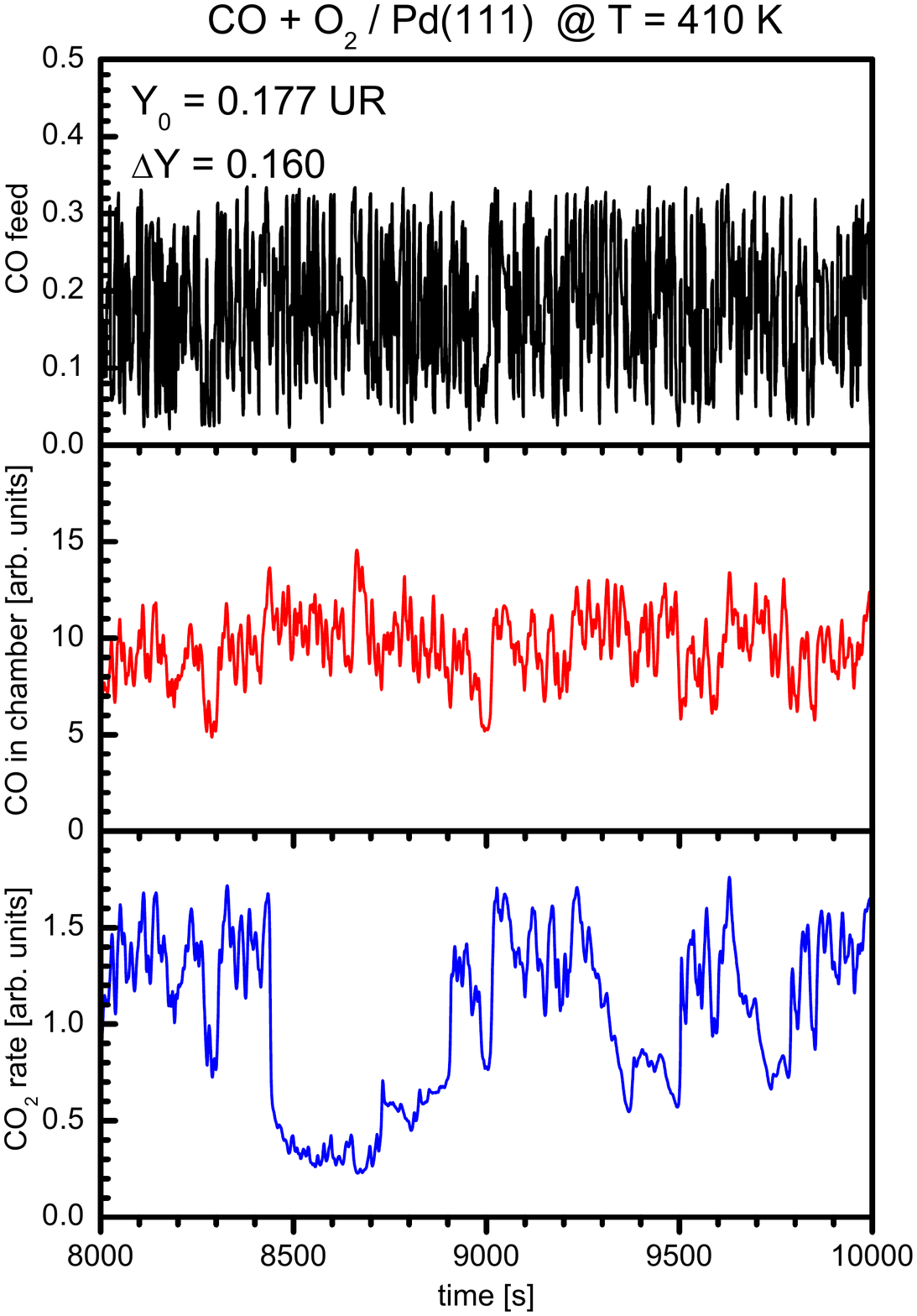}
\bigskip
\end{center}
\caption{As in figure \ref{fig:Y174_dY0000}, but for $Y_0 = 0.177$ and $\Delta Y = 0.160$ from Fig.~\ref{fig:Y177}}
\label{fig:Y177_dY0160}
\end{figure}

In Figs.~\ref{fig:Y174_dY0000}, \ref{fig:Y174_dY0006}, \ref{fig:Y174_dY0150} and \ref{fig:Y177_dY0160}, characteristic periods of experiments shown in previous figures are enlarged. For each experiment and from top to bottom, the CO feed (MFC reading), CO in chamber (QMS amu 28 signal) and $\COtwo$ rate (QMS amu 44 signal) are presented.
The top panel of Fig.~\ref{fig:Y174_dY0000} gives a good impression of the real fluctuations induced by intrinsic noise in the experimental setup used here, since CO feed is the reading from the MFC before smoothing by the capillary. The middle and bottom panels are QMS recordings inside the vacuum system. As described in the introduction, fragmentation of $\COtwo$ molecules inside the QMS contribute to the signal amu 28. But as the CO signal increases while the $\COtwo$ signal is decreasing during the transition starting at 33500 s, not the fragmentation but the non-consumption is dominating and the QMS reading for amu 28 represents the partial pressure of CO in the chamber.
For slightly larger noise $\Delta Y = 0.006$ the same behavior could be observed for the dip around 33500 s in Fig.~\ref{fig:Y174_dY0006}.
This global effect of the overall $\COtwo$ rate on the CO concentration will be addressed in the next section.

Figure \ref{fig:Y174_dY0150} shows an experiment with large noise $\Delta Y = 0.150$. Comparing CO in chamber (middle) and $\COtwo$ rate (bottom), it could be seen that both signals are \emph{in-phase} for the upper rate, check peaks at 17650 s, 17850 s, 18450 s and 18700 s. While at lower rate (during the switching in the interval 18000 s - 18300 s) the signals are in \emph{anti-phase}, check CO dip and $\COtwo$ peak at 18250 s. Comparing CO feed and CO in chamber it becomes clear that both are in-phase all the time as expected.
It is known that for CO oxidation on surfaces in UHV that in-phase behavior means predominantly oxygen covered and anti-phase behavior means predominantly CO covered.

These observations are also valid for a slightly higher $Y_0$ and $\Delta Y$, see Fig.~\ref{fig:Y177} for the recorded signal of an experimental run set with $Y_0 = 0.177$ and $\Delta Y = 0.160$ from 8000 s to 10000 s. 
These features of the induced fluctuations find a simple explanation in the slopes of the UR and LR branches of stable states (see right-upper panel in Fig.~\ref{fig:hysteresis}).
A noisy molar fraction effectively `rocks' the state of the system left-to-right in the bistable plot,
inducing in-phase or anti-phase fluctuations in $\COtwo$ rate depending on whether the slope of the branch is positive or negative respectively.

A more detailed examination of Figs.~\ref{fig:Y174_dY0150} and \ref{fig:Y177_dY0160} shows that for large noise, each of the transitions can be explained by the signal CO in chamber crossing certain thresholds.
Although these experiments do not provide images of the reaction on the surface, one of the main assumptions of the model presented in the next section is
that the thresholds can be traced back to the points where the UR and LR states become unstable.

\begin{figure}
\begin{center}
\includegraphics[width=0.5\textwidth]{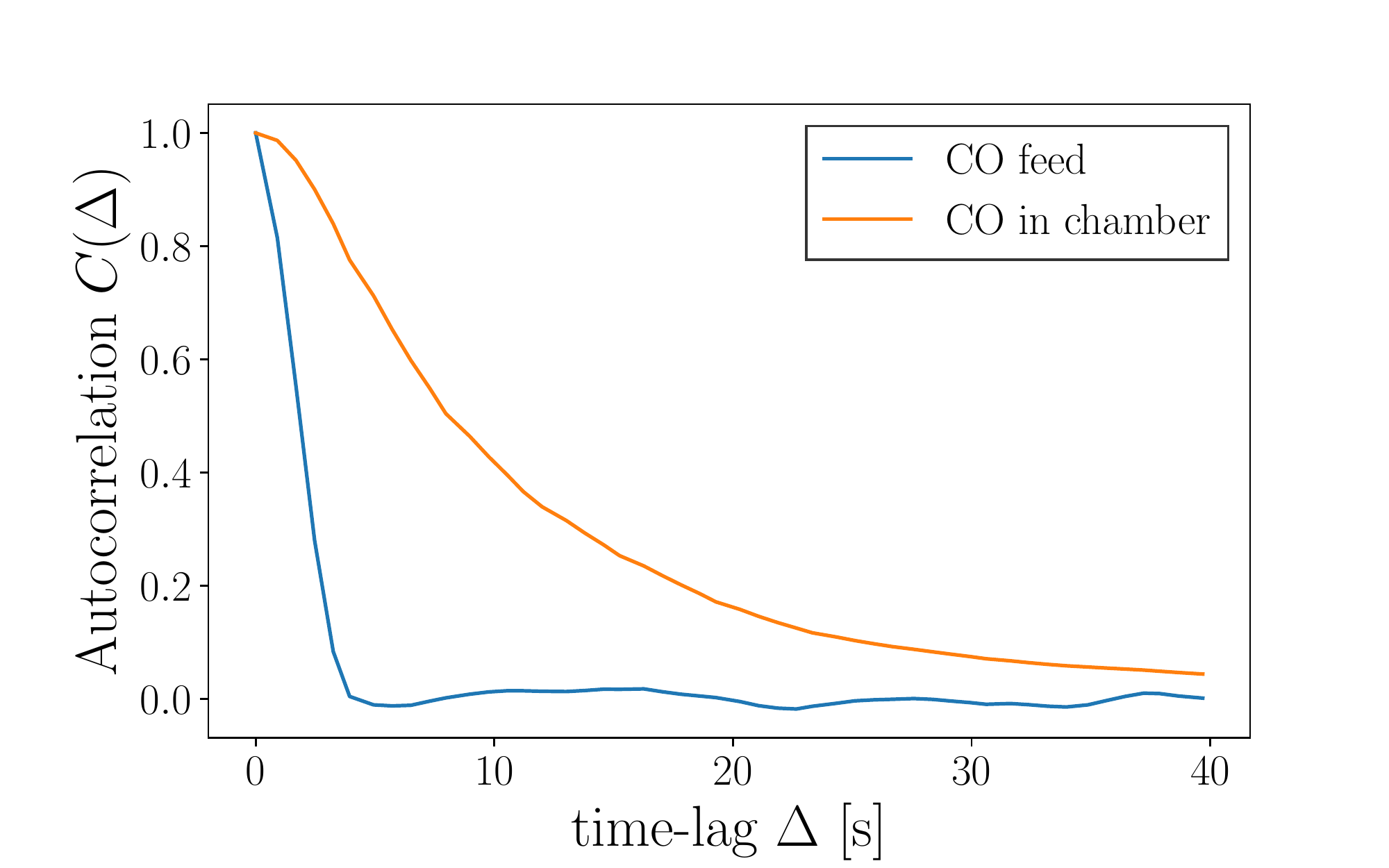}
\bigskip
\end{center}
\caption{Normalized autocorrelation functions of random signal CO feed that controls the $\CO/\Otwo$ ratio, and the experimentally observed CO concentration inside the chamber.
Although CO feed is the main driver of CO in chamber (see for instance Figs.~\ref{fig:Y174_dY0150} and \ref{fig:Y177_dY0160}),
the latter has a much longer characteristic time.}
\label{fig:autocorr}
\end{figure}

Another useful observation from Figs.~\ref{fig:Y174_dY0150} and \ref{fig:Y177_dY0160}
is the connection between the fluctuations of the signals CO feed and CO in chamber.
Ignoring the slow boosts in CO in chamber during the LR intervals,
it is possible to appreciate how CO in chamber follows the overall changes of the CO feed signal
but filters out the high frequency components.
A comparison of the normalized autocorrelations of these signals, in Fig.~\ref{fig:autocorr},
reveals that the characteristic time of CO feed signals is dictated by $t_\mathrm{step}=3$ s,
and that the characteristic time of CO in chamber $\tau \approx 15$ or $20$ s depends on the experimental setup
of the UHV chamber.



\section{Model and simulations}

As in other works studying the oxidation of CO on Platinum-group metals,
the starting point of the modeling will be
the coverages of CO and O on the surface.
Changes of these coverages reflect elementary processes such as the adsorption of a CO molecule or
the reaction between a CO molecule and a neighboring O atom
adsorbed on the crystal surface.
But the macroscopic scale of the Palladium crystal and the ultra high vacuum conditions allow
a continuous description of these CO and O coverages as the variables 
$u(x,t)$ and $v(x,t)$ respectively.

These coverages satisfy:
\begin{equation}
u \ge 0 ~,~ v \ge 0 ~,~ u+v \le 1 ~,
\end{equation}
and obey partial differential equations.
Here for simplicity we work in one spatial dimension so $x \in (0,L)$.

The dynamical equations for $u$ and $v$ are based on the Langmuir-Hinshelwood mechanism (compare e.g. Ref.~\onlinecite{KWK09}).
For the CO coverage (including diffusion, adsorption, desorption and reaction):
\begin{equation}
\frac{\partial u}{\partial t} = \nabla D_u(v) \nabla u + s_\CO \Phi_\CO (1-u)-k_\mathrm{des} u - k_\mathrm{rea} u v \label{dyneqn1}
\end{equation}
In this equation one can use coverage-dependent diffusion for CO \cite{CKW14}:
\begin{equation}
D_u(v) \mydef D_u^0 (1-v/v_\mathrm{max})^2 ~. \label{cdd}
\end{equation}
but the effect is minor in the context of noise.
There are other models that have been proposed for the diffusive transport \cite{VRN92,TE98a}
but we are using here the simplest form that is consistent with the observed phenomena.

For the O coverage (including diffusion, adsorption, and reaction):
\begin{equation}
\frac{\partial v}{\partial t} = D_v \nabla^2 v + 2 s_{\Otwo} \Phi_{\Otwo} (1-u-v)^{2} - k_\mathrm{rea} u v \label{dyneqn2}
\end{equation}

A more detailed justification of these terms, as well as values for the coefficients based on experiments, can be found in Ref.~\onlinecite{KWK09}.
Table \ref{thetable} presents a summary of the parameters used in simulations.
Adsorption terms are proportional to the gas concentrations $\Phi_\CO(t)$
(`CO in chamber' in Figs.~\ref{fig:Y174_dY0000}, \ref{fig:Y174_dY0006}, \ref{fig:Y174_dY0150}, and \ref{fig:Y177_dY0160})
and $\Phi_{\Otwo}(t)$
i.e. the number of $\CO$ and $\Otwo$ molecules hitting the Pd surface respectively.

\begin{table}
\caption{Parameters used in simulations.}
\begin{tabular}{cc}
\hline\hline
Parameter & Value \\
\hline
ML & $1.56\times 10^{15}\ \mathrm{cm^{-2}}$ \\
$s_\CO$ & $0.8$ \\
$s_{\Otwo}$ & $0.5$ \\
$k_\mathrm{des}$ & $\nu_\mathrm{des} \exp(E_\mathrm{des}/kT)$ \\
$E_\mathrm{des}$ & $125\ \mathrm{kJ/mol}$ \\
$\nu_\mathrm{des}$ & $5\times 10^{13}\ \mathrm{s^{-1}}$ \\
$k_\mathrm{rea}$ & $\nu_\mathrm{rea} \exp(E_\mathrm{rea}/kT)$ \\
$E_\mathrm{rea}$ & $53\ \mathrm{kJ/mol}$ \\
$\nu_\mathrm{rea}$ & $10^{7}\ \mathrm{(ML\ s)^{-1}}$ \\
$k$ & $0.008314472\ \mathrm{kJ/mol/K}$ \\
$T$ & $410\ \mathrm{K}$ \\
$\Phi$ & $1\ \mathrm{ML\ s^{-1}}$ \\
$D_v/D_u^0$ & $0.1$ \\
$v_\mathrm{max}$ & $0.4$ \\
$g$ & $0.5$ \\
$\tau$ & $20\ \mathrm{s}$ \\
\hline\hline
\end{tabular}
\label{thetable}
\end{table}

Desorption of oxygen is negligible and thus not included in the model.
Parameters such as $k_\mathrm{des}$ and $k_\mathrm{rea}$ depend on temperature,
but here they are assumed constant.

The external noise is imposed through a sequence of random numbers $Y(t)$ 
that are changed every $t_\mathrm{step}=3$ seconds and
drive the two mass flow controllers in the following form:
the concentration of CO that is fed into the chamber is $\Phi Y$ (`CO feed' in Figs.~\ref{fig:Y174_dY0000},\ref{fig:Y174_dY0006},\ref{fig:Y174_dY0150}, and \ref{fig:Y177_dY0160}),
and the concentration of $\Otwo$ that is fed into the chamber is $\Phi (1-Y)$, so
the sum of these two quantities must be at all times equal to the total flux of the feed gas $\Phi$.

Now the finite volume of the chamber and the finite power of the pumping systems induce two effects.
First, they prevent infinitely fast changes in the concentrations
$\Phi_\CO(t)$ and $\Phi_{\Otwo}(t)$.
Second, adsorption and desorption of CO and $\Otwo$ influence the number of molecules that are available for adsorption.
A basic model of this `global coupling', is written as two ordinary differential equations:
\begin{align}
\frac{d\Phi_\CO}{dt} = & \frac{1}{\tau} \left(\Phi Y - g \overline{(s_\CO \Phi_\CO (1-u)-k_\mathrm{des} u)} - \Phi_\CO(t) \right) \label{globalcouplin1}\\
\frac{d\Phi_{\Otwo}}{dt} = & \frac{1}{\tau} \left(\Phi (1-Y) - g \overline{(s_{\Otwo} \Phi_{\Otwo} (1-u-v)^2)} - \Phi_{\Otwo}(t) \right) \label{globalcouplin2}
\end{align}
where the horizontal lines mean spatial averages over the whole surface.
The factor $g/\tau$ is proportional to surface area and inversely proportional to chamber volume.
A detailed analysis of the experimental observations,
summarized in Figs.~\ref{fig:Y174_dY0000},\ref{fig:Y174_dY0006},\ref{fig:Y174_dY0150},\ref{fig:Y177_dY0160}, and \ref{fig:autocorr},
indicates that $\tau$ is of the order of 20 seconds,
and that the $g/\tau$, that best approximates the enhanced CO concentration in the chamber
when the state reaches LR, is $g/\tau \approx 0.025$.

A direct effect of global coupling is that the sum of the two gaseous concentrations
$\Phi_\CO(t)+\Phi_{\Otwo}(t)$ is no longer constant and can have a destabilizing effect.



\begin{figure}
\begin{center}
\includegraphics[width=0.5\textwidth]{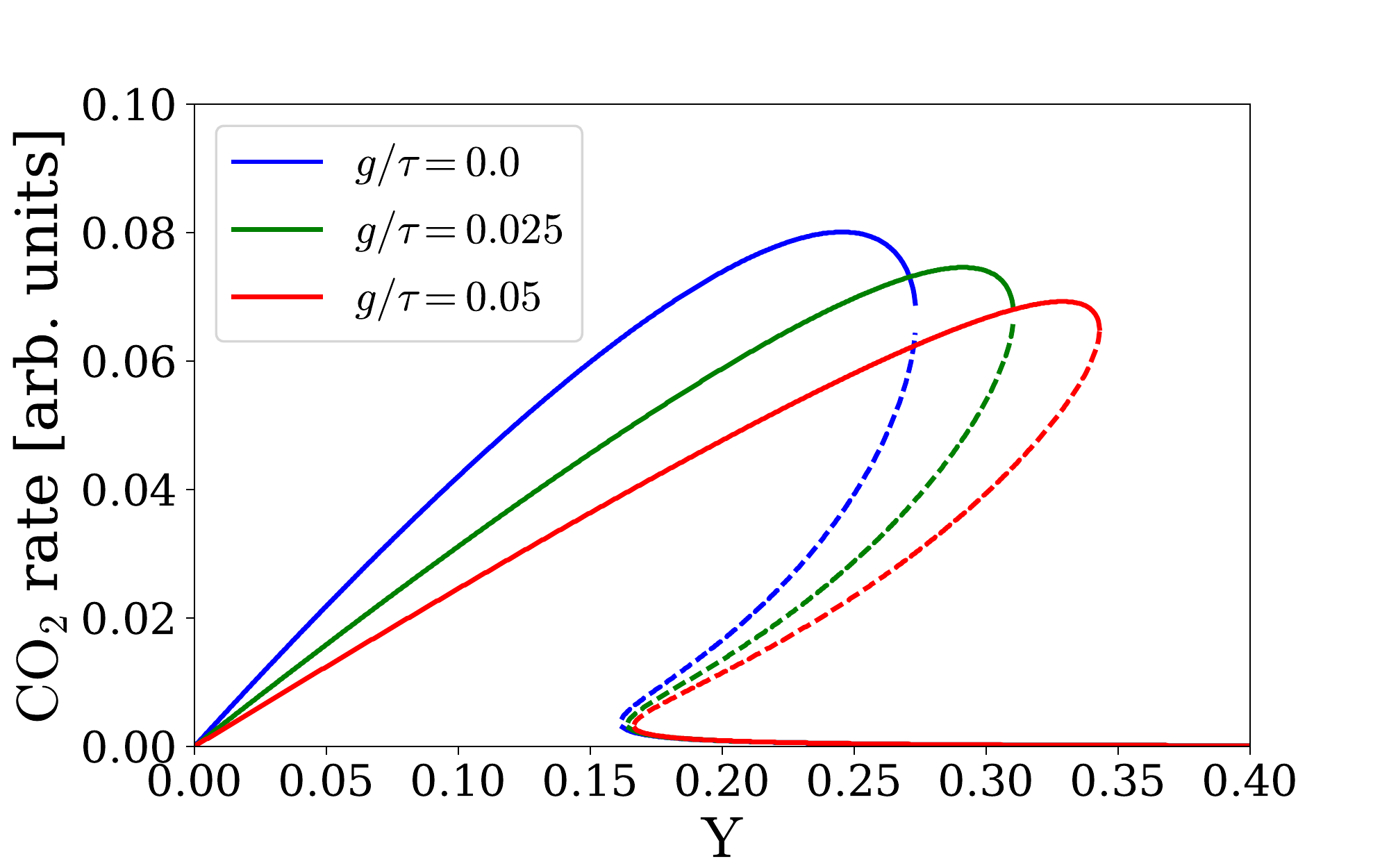}
\end{center}
\caption{Global coupling, modeled by Eqns.~(\ref{globalcouplin1},\ref{globalcouplin2}),
addresses the finite volume of the chamber and the finite surface of the crystal.
When constant and homogeneous coverages are considered,
non-zero values of $g$ make the UR state more stable and enlarge the region of bistability.
Here the state of the system is measured by the $\COtwo$ reaction rate equal to $k_\mathrm{rea} uv$.
}
\label{fig:bistability}
\end{figure}

Global coupling can potentially modify the basic theoretical picture of the reaction,
shifting stable states or explaining additional instabilities\cite{LZ93,BHE94}.
Fig.~\ref{fig:bistability} shows the effect of the parameter $g$ in the equilibrium states
of the proposed model defined in Eqns.~(\ref{dyneqn1}--\ref{globalcouplin2}).
As a larger value of $g$ (a larger surface to volume quotient) is used,
the location of the right-most fold ($Y_h$) is shifted to the right 
and the UR branch increases its range of existence.
Only constant and homogeneous coverages were considered in the figure.

More interesting phenomena may take place when considering heterogeneous coverages.
For instance, when a LR island is growing and advancing on a UR background,
the concentration $\Phi_{\CO}(t)$ grows and the `additional' stability of the UR state decreases,
fostering an acceleration of the growth of the LR island, until the whole UR state becomes unstable and
a global transition to the LR takes place.
The enhanced $\Phi_{\CO}(t)$ furthers the stability of LR to the point of making the back transition to UR impossible.

\begin{figure}
\begin{center}
\includegraphics[width=0.5\textwidth]{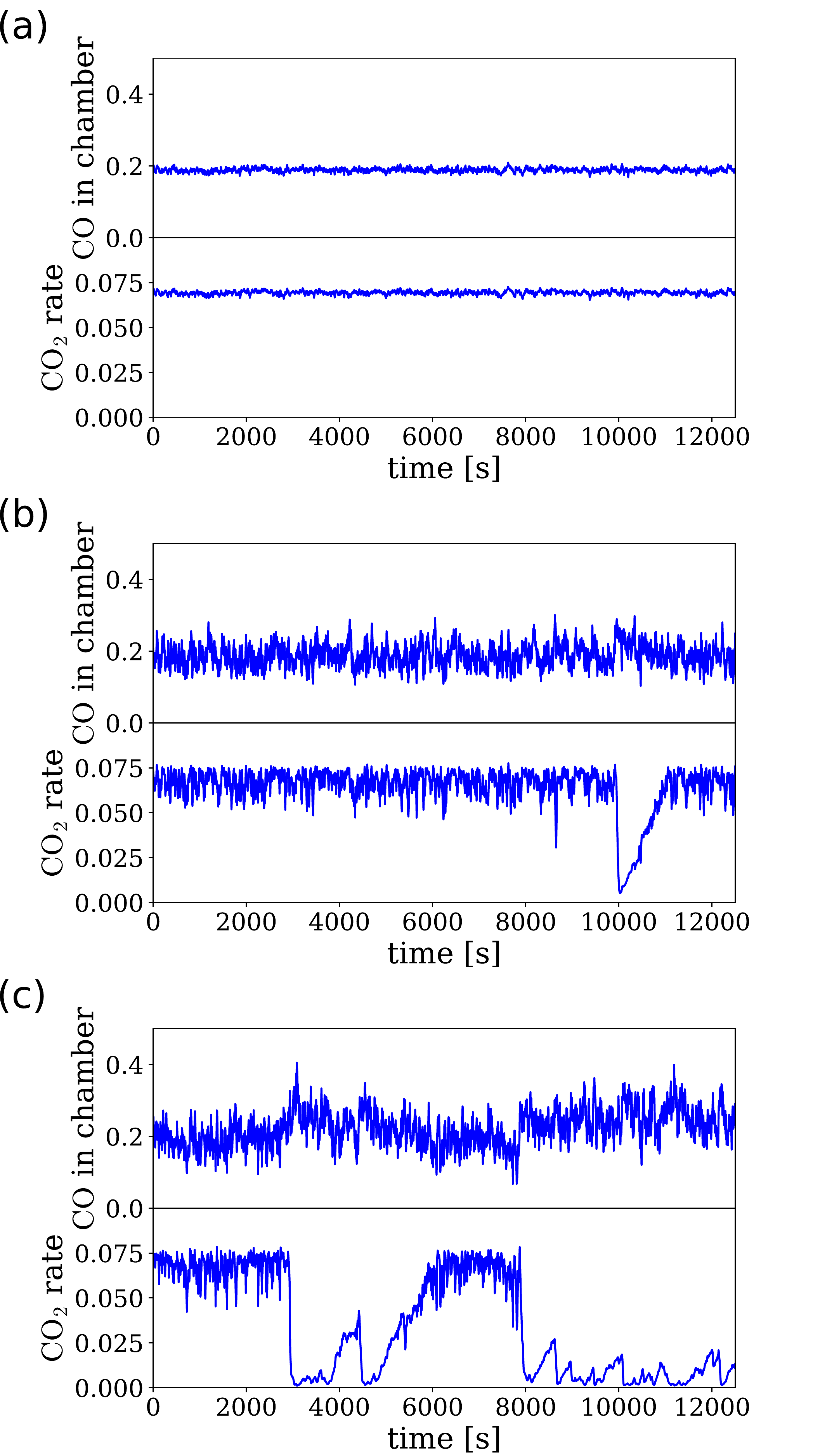}
\bigskip
\end{center}
\caption{Summary of behaviors obtained from the model, for $Y_0=0.25$ and three different noise intensities:
(a) small noise $\Delta Y=0.02$; (b) medium noise $\Delta Y=0.11$; (c) large noise $\Delta Y=0.13$.
For each simulation, the plots show the evolution of $\COtwo$ rate and the CO concentration inside the chamber.}
\label{fig:model1}
\end{figure}

\begin{figure}
\begin{center}
\includegraphics[width=0.5\textwidth]{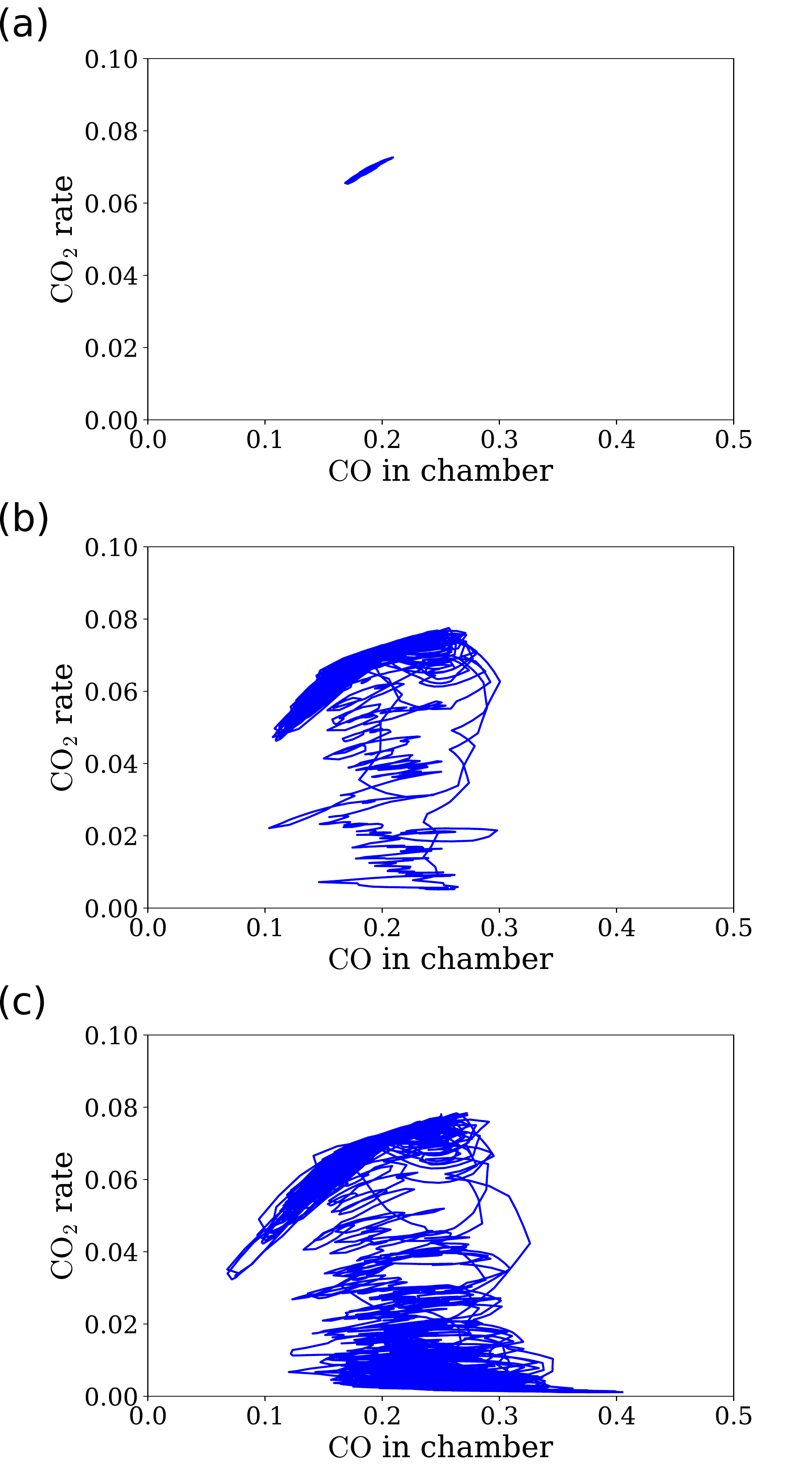}
%
%
\end{center}
\caption{Summary of behaviors obtained from the model, for $Y_0=0.25$ and three different noise intensities:
(a) small noise $\Delta Y=0.02$; (b) medium noise $\Delta Y=0.11$; (c) large noise $\Delta Y=0.13$.
For each simulation, the plots show $\COtwo$ rate versus the CO concentration inside the chamber.}
\label{fig:model2}
\end{figure}

In Figs.~\ref{fig:model1} and \ref{fig:model2} we present typical results of the proposed
model defined in Eqns.~(\ref{dyneqn1}--\ref{globalcouplin2})
for three different noise intensities $\Delta Y$.
To facilitate comparison with the experimental plots, we use the following variables:
\begin{align*} 
\CO \text{~inside chamber} &= \Phi_{\CO}(t) ~, \\
\COtwo \text{~reaction rate} &= \int_0^L k_\mathrm{rea} u(x,t) v(x,t)\ dx ~.
\end{align*} 
Simulations were performed using a second-order operator splitting made up of
a semi-implicit (Crank-Nicolson) method for the spatial derivatives,
and a second-order Runge-Kutta method for the reaction terms.
Fixed boundary conditions were used at the two endpoints of a one-dimensional domain of length $L=50$ (arbitrary spatial unit):
\begin{align*} 
u(0,t) = u_\mathrm{LR} ~,~ & v(0,t) = v_\mathrm{LR} ~, \text{~and~~} \\
u(L,t) = u_\mathrm{UR} ~,~ & v(L,t) = v_\mathrm{UR} \label{bc} ~.
\end{align*} 

For small noise intensity $\Delta Y = 0.02$, as depicted in Figs.~\ref{fig:model1}(a) and \ref{fig:model2}(a),
both signals mimic a low-pass filtered version of $Y(t)$. No transitions are observed over a time period of $10000$ s.
The in-phase fluctuations of the $\COtwo$ reaction rate can be understood in Fig.~\ref{fig:model2}(a)
as a remnant of the characteristic shape of the UR branch (see for instance the upper panel of Fig.~\ref{fig:hysteresis} or Fig.~\ref{fig:bistability}).


For medium noise intensity $\Delta Y = 0.11$, shown in Figs.~\ref{fig:model1}(b) and \ref{fig:model2}(b), the surface reaction becomes dominated by noise,
exhibiting fluctuations and bursts around the UR state, that are not able to develop as full transitions to the LR state.

Figs.~\ref{fig:model1}(c) and \ref{fig:model2}(c) show the dynamics induced by large noise intensity $\Delta Y = 0.13$.
Both transitions, from UR to LR and from LR to UR, are now observed.
The $\COtwo$ reaction rate vs. $\CO$ inside the chamber shows a noisy trajectory that closely follows
the hysteresis loop made up of the UR and LR branches.
The basic observations about the phase of the fluctuations of the $\COtwo$ reaction rate in the UR and LR states still apply.
Although the transitions follow a \emph{clockwise} direction in Fig.~\ref{fig:model2}(c) most of the time,
several `failed' or incomplete transitions occupy the central part of the loop.


Overall we can appreciate how the experimental results and the numerical results obtained from
our model with global coupling agree at the qualitative level, for the three different regimes.
For the noise-free case, the model is so far unable to replicate the observed transitions.

\begin{figure}
\begin{center}
\includegraphics[width=0.5\textwidth]{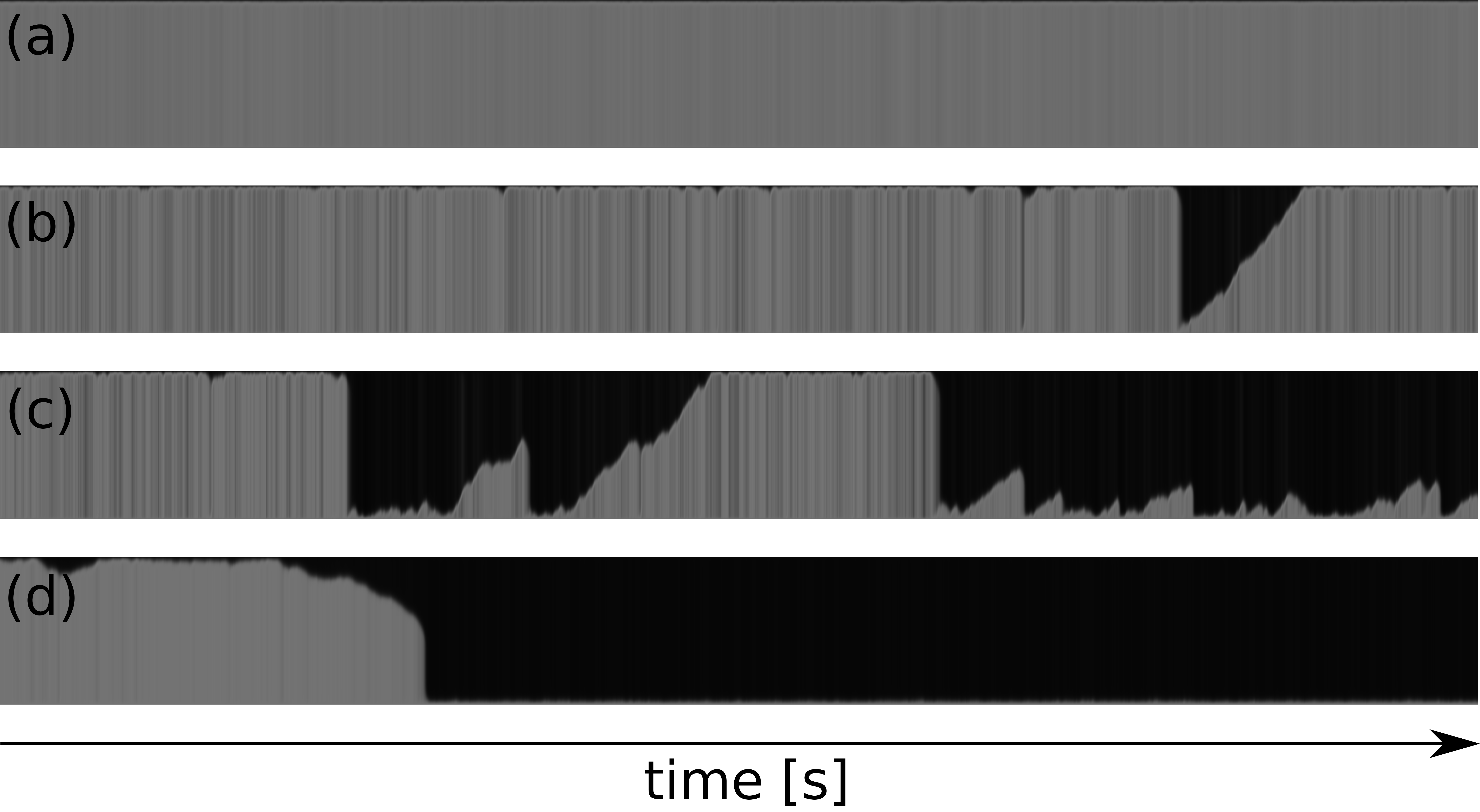}
%
%
%
\end{center}
\caption{Summary of behaviors obtained from the model, for four different combinations of
average molar fraction and noise strength:
(a) $Y_0=0.25, \Delta Y=0.02$; (b) $Y_0=0.25, \Delta Y=0.11$; (c) $Y_0=0.25, \Delta Y=0.13$; (d) $Y_0=0.29, \Delta Y=0.04$.
For each simulation, the grayscale indicates whether the local state of the coverage is closer to
UR (white) or LR (black). For each plot, horizontal axis is time as it grows to the right,
and vertical axis is spatial coordinate.}
\label{fig:model3}
\end{figure}

In Fig.~\ref{fig:model3} we include some representative spacetime plots that intend to mimic experimentally obtained PEEM plots\cite{WKBSKB10}.
The grayscale is adjusted so UR state appears white and LR black.
The upper boundary ($x=0$) is set to LR, and the bottom boundary ($x=L$) to UR.
For the three subfigures $Y_0$ was chosen so UR is more stable than LR and UR regions would spontaneously grow.
This explain the asymmetry in the figures: UR to LR transitions are fast and global;
while LR to UR transitions are slow because they are mediated by fronts (at least initially).

The last figure, Fig.~\ref{fig:model3}(d), shows how for a value $Y_0$ closer to the critical point $Y_h$
and small noise, an accelerated and irreversible transition from UR to LR is observed.
The enhanced concentration $\Phi_{\CO}$ associated to the LR state makes the opposite transition impossible
for such a weak noise, a direct consequence of the global coupling introduced into the model.
This modelling result is clearly connected with the experimental results in Fig.~\ref{fig:Y177} (two uppermost plots)
showing CO poisoning.

It is also interesting to compare the results of the heterogeneous model and those of an homogeneous model
i.e. that considers only homogeneous coverages $u(t)$ and $v(t)$.
In our simulations the homogeneous model showed far fewer transitions within a finite time.
Although other factors such as the role of surface defects are ignored in the present study
there are indications that global coupling and certain boundary conditions may facilitate transitions.

These observations apply to a large number of independent realizations of the noise signal $Y(t)$ over a time period of $10000$ s.
We expect the number of noise-induced transitions to increase when considering a larger $\Delta Y$ or longer periods.
In the limit of infinitely long time spans and for nonzero $\Delta Y$ the model should show both transitions an infinite number of times,
in principle.

\section{Conclusions}

Several previous studies have considered the influence of noisy conditions on the CO oxidation on Platinum group metals under ultra high vacuum conditions.
In this article we focused on Palladium(111) near $Y_{h}$, the right limit of the bistable range, and took advantage of detailed measurements of gaseous concentrations inside the vacuum chamber. 

Our experimental results showed that there are several regimes depending basically on noise intensity.
In the absence of external noise (when only intrinsic noise is present) and after thousands of seconds
the state of the system (as monitored by the reaction rate) exhibits a global transition from the UR state to the LR state
(spontaneous CO poisoning)
from where it remains thereafter.
These transitions induce an increase in CO concentration inside the chamber.
For small noise, transitions are suppressed.
For medium noise, large bursts are observed.
For large noise, both transitions are observed.

The model that was presented here is based on the Langmuir-Hinshelwood mechanism.
It takes into consideration the external noise as well as the influence of the surface processes
on the gas concentrations inside the chamber.
The dynamics obtained from the model reproduced at the qualitative level the basic findings of the experiments:
the existence of three regimes of noise-induced transitions;
the enhancement of CO inside the chamber during LR periods; and
minute details of the fluctuations of $\COtwo$ reaction rate during the UR and LR periods.

The present work could be expanded with a thorough study of the dynamics of the two-dimensional coverages
and the influence of boundary conditions and heterogeneities,
both in modeling and in experiments using imaging techniques such as PEEM or EMSI.
As it can be anticipated from the analysis of global coupling,
the acceleration effect of fronts should be even more noticeable in two dimensions.
In addition, fronts in two dimensions suffer from instabilities that are not relevant in one dimension.
These instabilities can have effects on the features of the transitions in both directions.
The instabilities may also explain the reduced reaction rates observed after `failed' transitions to the lower rate.

\section*{Acknowledgments}

The authors thank J\"urgen K\"uppers for enabling these research opportunities,
and Stefan Karpitschka for the software used and for basic studies without noise.
J.C. thanks the financial support of FONDECYT Project 1170460.



\end{document}